\documentclass[aps,10pt,twocolumn,floatfix,amsmath,amssymb,prl,longbibliography,raggedbottom]{revtex4-1}
\usepackage{graphicx}
\usepackage{color}

\usepackage{bm}
\usepackage{amsmath}

\newcommand{\red}[1]{#1}

\begin{document}

\title{
Random-matrix perspective on many-body entanglement\\
with a finite localization length}

\author{Marcin Szyniszewski}
\affiliation{Department of Physics, Lancaster University, Lancaster, LA1 4YB, United Kingdom}
\author{Henning Schomerus}
\affiliation{Department of Physics, Lancaster University, Lancaster, LA1 4YB, United Kingdom}

\begin{abstract}
We provide a simple and predictive random-matrix framework that naturally generalizes Page's law for ergodic many-body systems by incorporating a finite entanglement localization length. By comparing a highly structured one-dimensional  model to a completely unstructured model and a physical system, we uncover a remarkable degree of universality, suggesting that the effective localization length is a universal combination of model parameters up until it drops down to the microscopic scale. 
\end{abstract}

\maketitle

In this paper we present a generalization of Page's law \cite{Page1993}---central to the statistical description of entanglement in completely ergodic many-body systems---so that it incorporates a finite entanglement length scale, designed to represent an effective localization length in a many-body localized system \cite{Basko2006,Altman2015,Nandkishore2015,Abanin2017,Alet2018}. Page's law is based on the simple assumption that a typical ergodic  many-body eigenstate $|\psi \rangle$ constitutes a random Fock-space vector with independent identically distributed Gaussian entries $\psi_m=\langle m|\psi \rangle$. Bipartitioning the system as a tensor product $| m\rangle=| ab\rangle$, with indices $a=1,2,3,\ldots, M_\mathsf{A}$ for a subsystem \textsf{A} and $b=1,2,3,\ldots, M_\mathsf{B}$ for its complement \textsf{B}, the reduced density matrix $\rho_{aa'}^\mathsf{A|B}$ of a subsystem \textsf{A} can be reinterpreted as a matrix product,
\begin{equation}
\rho^\mathsf{A|B}=\frac{V V^\dagger}{\mathrm{tr}\,V V^\dagger},
\end{equation}
where $V_{ab}=\langle ab|\psi \rangle$ is a random $M_\mathsf{A}\times M_\mathsf{B}$ Gaussian matrix. This ties the description to the celebrated Wishart ensemble of random matrix theory---the inaugural ensemble of random matrix theory in the history of science \cite{Wishart1928}, which is based on completely positive Hermitian matrices of the form $V V^\dagger$. Applying these arguments, Page then arrived at the prediction
\begin{align}
S(\mathsf{A}|\mathsf{B}) = -\mathrm{tr}\,(\rho^\mathsf{A|B} \ln \rho^\mathsf{A|B})
= \ln M_\mathsf{A}-\frac{M_\mathsf{A}}{2M_\mathsf{B}}
\label{eq:page}
\end{align}
for the ensemble-averaged bipartite von Neumann entanglement entropy, assuming
$1\ll M_\mathsf{A} \leq M_\mathsf{B}$ \footnote{In this limit, the result is
independent of the universality class of random-matrix theory.}. This prediction
serves as an important benchmark to detect deviations from ergodic many-body
behavior, including signatures of many-body localization and topological states
\cite{Bauer2013, Kjaell2014, Vidmar2017, LeBlond2019, Khemani2017}.

Here, we present a simple and predictive statistical framework that accurately
\emph{captures} these deviations, and covers the distance from entirely ergodic
behavior to the strongly many-body localized regime. This gives very direct and
specific insights into a transition that so far has been addressed mainly
through insightful perturbative strong-disorder renormalization schemes
\cite{Potter2015, Vosk2015, Zhang2016, Dumitrescu2017, Thiery2018,
Goremykina2019}. We first give a simple motivation and description
of the framework and analyze its main features, amongst which is a surprising
degree of universality with regards to both the microscopic parameters as well
as the local structure of the random-matrix model. We then demonstrate its
predictive power in comparison with a paradigmatic spin-chain model. Finally, we
discuss the framework from the general perspective of matrix-product states.

\begin{figure}[t]
  \includegraphics[width=\columnwidth]{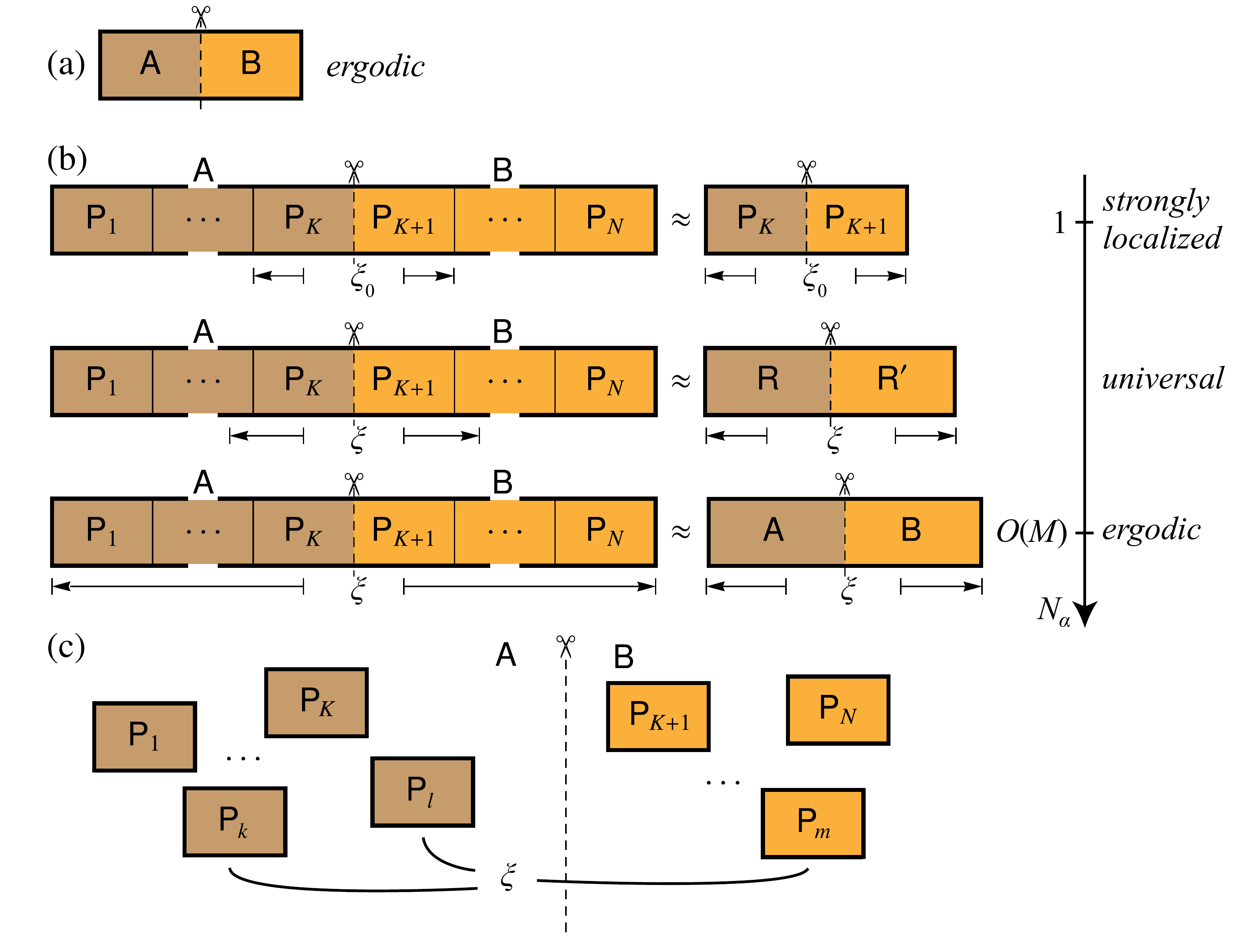}
  \caption{(a) Page's law \eqref{eq:page} describes the  entanglement in an ergodic system partitioned into parts \textsf{A} and \textsf{B}.
  (b) The highly structured model \eqref{eq:model} describes the transition from a strongly localized regime, where the bipartitioned system can be effectively reduced to two small ergodic patches next to the partition point, over a universal regime with a finite localization length \eqref{eq:xieff}, to the ergodic case where Page's law is recovered. (c) In the completely unstructured variant \eqref{eq:model2}, a similar transition occurs but the ergodic regions may be interpreted as non-contiguous.
  \label{fig1}}
\end{figure}

\begin{figure}[t]
  \includegraphics[width=\linewidth]{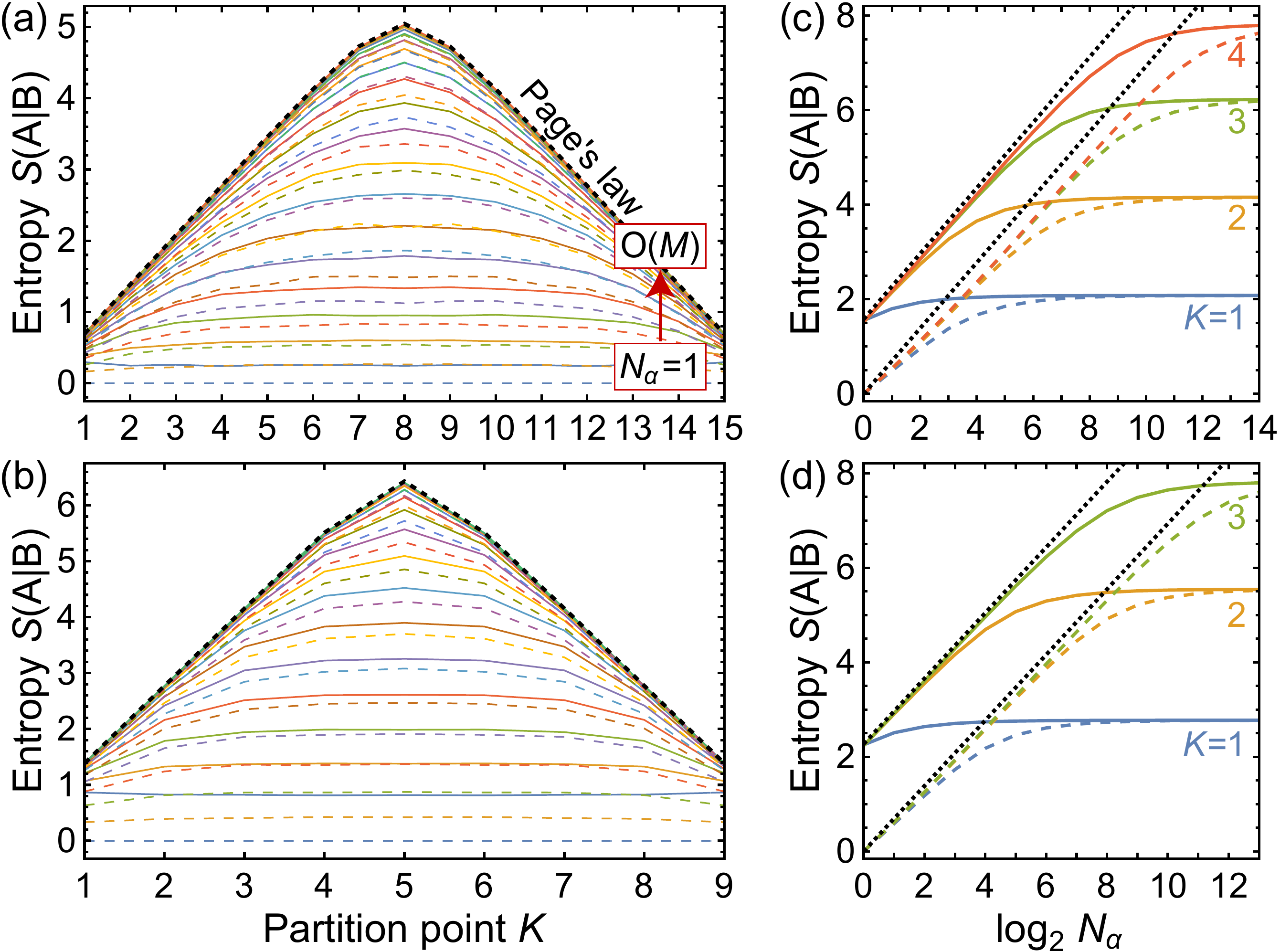}
  \caption{Bipartite entanglement entropy $S(\mathsf{A}|\mathsf{B})$ as predicted by the structured model \eqref{eq:model} (solid curves) and the unstructured model \eqref{eq:model2} (dashed curves), for different partitions $\mathsf{A}|\mathsf{B}=\mathsf{P}_1\ldots \mathsf{P}_K | \mathsf{P}_{K+1}\ldots$, where each patch has $2^{N_0}$ internal states.
  In (a) the system is made of 16 patches of size $N_0=1$, in (b) of 10 patches of size $N_0=2$,
  in (c) of 8 patches of size $N_0=3$, and in (d) of 6 patches of size $N_0=4$.
  Panels (a,b) highlight the dependence with the partition point for different numbers of superimposed states $N_\alpha=2^m$, $m=0,1,2\ldots$, where the results for large $m$ approach the ergodic result from Page's law \eqref{eq:page} (thick dotted curve).
  Panels (c,d) highlight the dependence on $N_\alpha$, where the sloped dotted lines correspond to an ergodic system truncated to the effective localization length \eqref{eq:xieff}; for large $N_\alpha$ the curves level off at Page's law. We find excellent agreement with the predicted universal behavior, which sets in quickly for increasing patch and system size.
    \label{fig2}}
\end{figure}

\emph{Premise and background.---}To motivate our approach it is suggestive to declare Page's statistical assumptions as natural in the following sense: The matrix $V$ can be interpreted to capture the correlation amplitudes between the adjacent parts $\mathsf{A}$ and $\mathsf{B}$, in a statistical invariant way where for instance any independent superposition
$V=\sum_{\alpha=1}^{N_{\alpha}} V^{\alpha}$ of matrices from the same Gaussian ensemble delivers the same statistics
\footnote{Under very general conditions the central-limit theorem guarantees that even if the $V^{\alpha}$'s were to be taken from a highly structured distribution (but forming a complete basis in a statistical sense), the ergodic behavior would be recovered for large $N_{\alpha}$. This feature is exploited in the ergodic limit of our model.}.
In our generalization, the system is partitioned into a larger number of small ergodic patches $\mathsf{P}_1,\mathsf{P}_2,\mathsf{P}_3,\ldots$, which we take of identical dimensionality $M_0$, and the wave function takes the simple form
\begin{equation}
  \psi_{abcd\ldots} = \sum_{\alpha=1}^{N_\alpha} V^{1|2,\alpha}_{ab}
  V^{2|3,\alpha}_{bc} V^{3|4,\alpha}_{cd} \ldots .
\label{eq:model}
\end{equation}
The random Gaussian matrices $V^{k|k+1,\alpha}$ again describe the correlations
between neighboring ergodic patches, only that there are now many of these. This
defines the \emph{highly structured} variant of our model. Taking, in contrast,
the matrices $V$ as separable, we arrive at a \emph{completely unstructured}
model equivalent to a superposition of completely separable states
\begin{equation}
  \psi_{abcd\ldots} = \sum_{\alpha=1}^{N_\alpha} \chi^{1,\alpha}_{a}
  \chi^{2,\alpha}_{b} \chi^{3,\alpha}_{c} \chi^{4,\alpha}_{d} \ldots
\label{eq:model2}
\end{equation}
with random amplitudes $\chi$, which is agnostic about the ordering of the
patches and hence does not contain any information about geometric features,
such as dimensionality and boundary conditions. The interplay of model
\eqref{eq:model} and \eqref{eq:model2} defines our random matrix framework. In
both cases, the reduced density matrix for a bipartition $\mathsf{A}|\mathsf{B}
= \mathsf{P}_1\ldots \mathsf{P}_K|\mathsf{P}_{K+1}\mathsf{P}_{K+2}\ldots$ is
obtained by tracing out the sequence of patches
$\mathsf{P}_{K+1}\mathsf{P}_{K+2}\ldots$.

We will argue, and verify numerically, that this framework identifies key
entanglement characteristics of systems with a finite range of the entanglement,
subsumed into a \emph{universal} effective localization length $\xi$ that
combines the microscopic model parameters $M_0$ and $N_\alpha$ into one. The
universality is fully established in the mesoscopic regime, where the parts are
all small compared to the size of the bipartioned subsystems, and $N_\alpha$ is
moderately large, but in practice already holds well for $N_\alpha=O(1)$. In
particular, in comparison to physical models the framework turns out to be
remarkably predictive for the bipartite entanglement entropy at different system
sizes and choice of bipartition \cite{Kaufman2016, Vidmar2017, LeBlond2019,
Shtanko2020}. As this universality is observed also \emph{between} the two
variants of the model, we conjecture that it also extends to interpolating
scenarios, including multifractal cases \footnote{By modifying the matrices
$V^{k|k+1,\alpha}$ gradually it is in principle possible to interpolate between
models \eqref{eq:model} and \eqref{eq:model2}}.

The key idea of the model, namely partitioning the system into small patches
$\mathsf{P}_k$ of size below the universal localization length scale and then
considering superpositions to arrive at a universal mesoscopic regime, is shown
pictorially in Fig.~\ref{fig1}. A complementary approach has been taken before
by several groups \cite{Potter2015, Vosk2015, Zhang2016, Dumitrescu2017,
Thiery2018, Goremykina2019}, who set up insightful perturbative strong-disorder
renormalization schemes for many-body localized systems based on coupling
strengths and thermalization rates between coupled blocks. In contrast, our
statistical approach directly stipulates the wave functions of the composed
system. Conceptually, this wave-function centered construction starting from the
ergodic limit has its precedent in powerful approaches to single-particle
Anderson localization. In influential papers by Dorokhov, Mello, Pererya and
Kumar (DMPK) \cite{Dorokhov1982,Mello1988} it has been shown that Anderson
localization naturally arises from the multiple scattering in a chain of
individual weakly scattering components. Analogously, Iida, Weidenm{\"uller},
and Zuk (IWZ) \cite{Iida1990}, showed that the same universal behavior emerges
from the multiple scattering in a chain of individually strongly scattering
components. The DMPK model and the IWZ model both attain the same universal
thick-wire fixed point as the supersymmetric $\sigma$ model \cite{Efetov1996,
Beenakker1997}, which is governed by a single length scale, the single-particle
localization length, in accordance to the one-parameter scaling hypothesis
\cite{Abrahams1979}. The model and predictions presented in our work can
serve as a benchmark to establish to which extent an analogous form of
one-parameter scaling applies to many-body localization. For establishing a
gradually more ergodic behavior of wavefunctions by superpositions, a useful
references point is Berry's random wave model for quantum-chaotic systems
\cite{berry1977regular}, which we here effectively carry over to Fock space
based on wavefunctions equipped with a suitable local structure. With this
wavefunction-centered approach, the resulting framework complements the
application of random-matrix theory as a benchmark for energy-level statistics
\cite{Oganesyan2007, Altman2015, Nandkishore2015, Suntajs2019}. \red{Other
phenomenological approaches gained recently substantial interest, e.g. in
systems with imposed entanglement restrictions~\cite{Morampudi2020}, as they
provide simple yet predictive pictures of phenomena that are otherwise
difficult to describe microscopically.}

\emph{Key features.---}To identify the key features of the highly structured
model \eqref{eq:model} and the completely unstructured model \eqref{eq:model2},
Fig.~\ref{fig2} shows the bipartite entanglement entropy obtained in systems of
different length and patch size. For definiteness, we phrase the length scales
in the language of systems with $N$ spins, broken down into small patches of
length $N_0$ (thus $M_0=2^{N_0}$).

In the left panels, we vary the partition point while keeping $N_\alpha$ fixed.
For $N_\alpha=O(1)$, the entropy is small and independent of the partition
point, corresponding to a highly localized system. For increasing $N_\alpha$,
the entropy rises, and finally attains the ergodic result \eqref{eq:page} for
the \emph{complete} system, which now depends on the partition point.
Remarkably, the curves of both models match up closely in the crossover.

We will reveal below by statistical arguments that this amounts to a universal
behavior governed by a single parameter in each model, an effective localization
length
\begin{equation}
  \xi\sim\xi_0+2\log_2 N_\alpha.
\label{eq:xieff}
\end{equation}
In particular, for $N_\alpha=1$ the entropy in the structured model closely
conforms to the ergodic result \eqref{eq:page} for a \emph{reduced} system with
only two patches adjacent to the partition point, hence effective localization
length $\xi=\xi_0=2N_0$. In the unstructured model, the entropy vanishes in this
limit, so that $\xi_0=0$. Increasing $N_\alpha$ then amounts to gradually
increasing the effective range of ergodic behavior, with a universal scaling of the effective
localization length. This universal scaling is verified in the right panels,
where we keep the partition point fixed. We see that the universal scaling is
attained quickly for moderately large patch and system sizes.


\begin{figure}[t]
  \includegraphics[width=\columnwidth]{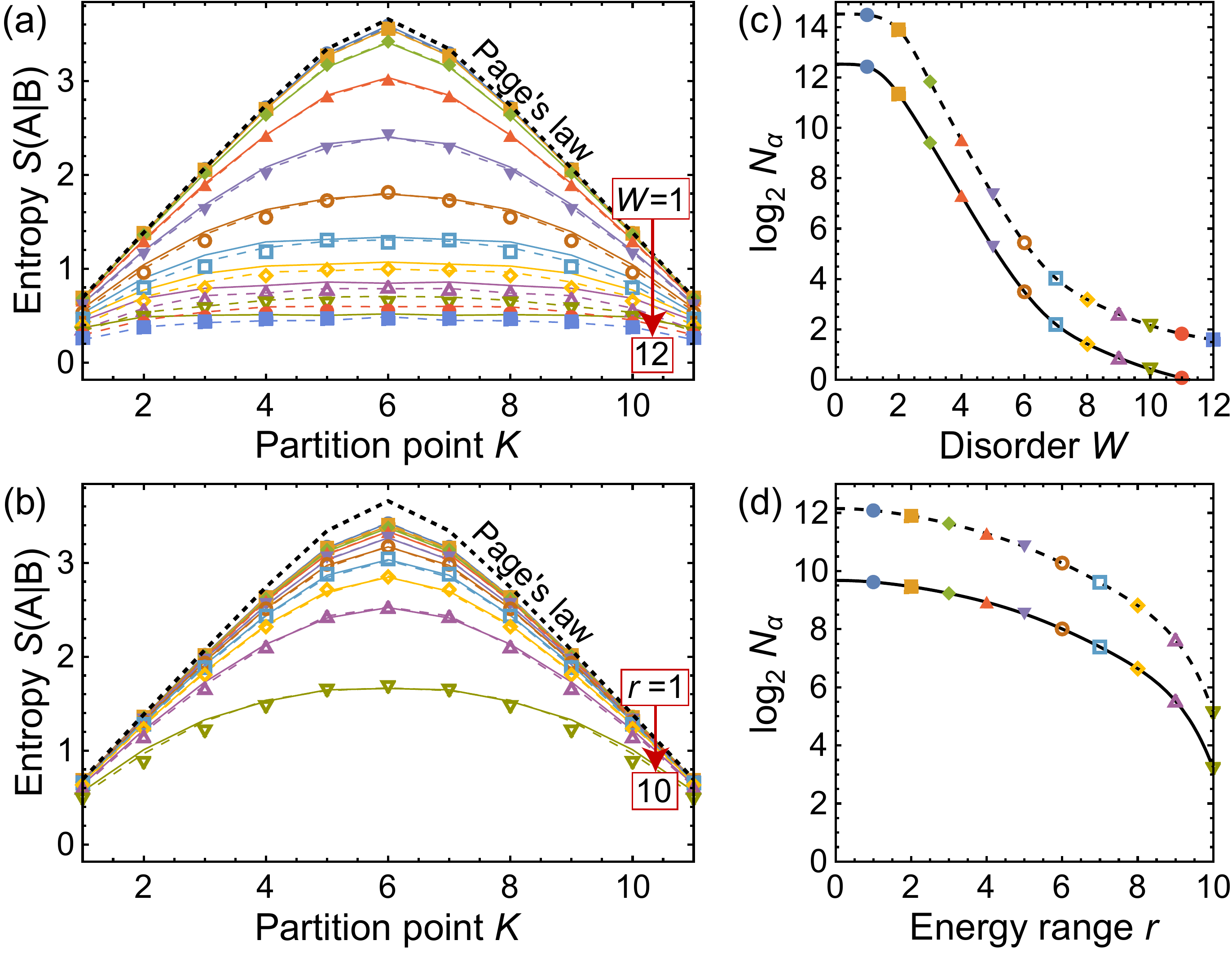}
  \caption{
  (a,b) The markers show the bipartite entanglement entropy in spin-chain
  \eqref{eq:ham} with $N=12$ sites as a function of the partition point $K$, for
  (a) different strengths of disorder $W=1,2,\ldots,12$ obtained from the 10\%
  of states closest to the band center, or (b) at fixed disorder strength $W=3$ with the states separated by energy into 10 groups ranging from the band center (where the entropy is large, range $r=1$) to
  the band edge (where it is small, $r=10$). The thick
  dashed curve indicates Page's law \eqref{eq:page} for the ergodic limit, while
  the thin solid and dashed curves show the corresponding predictions from the
  random-matrix models \eqref{eq:model} and \eqref{eq:model2} in analogy to
  Fig.~\ref{fig2}, with patch size $N_0=1$. In (c,d), the corresponding values
  of $\log_2 (N_\alpha)$ are plotted for the same system size as a function of 
  disorder strength or
  energy range, which delivers the effective localization length
  \eqref{eq:xieff}.
  \label{fig3}}
\end{figure}

\emph{Application to physical models.---}Below, we will give a detailed
statistical justification of this universal behavior. First, we describe how it
conforms and applies to concrete physical systems. This is illustrated in
Fig.~\ref{fig3}, where we provide a comparison to results for a spin chain with
Hamiltonian
\begin{eqnarray}
  \hat{H} & = & \sum_n\mathbf{h}_n\cdot\boldsymbol{\sigma}_n-\sum_n
  \boldsymbol{\sigma}_n\cdot \boldsymbol{\sigma}_{n+1},
  \label{eq:ham}
\end{eqnarray}
where $\boldsymbol{\sigma}_n$ is a vector of Pauli matrices on the $n$-th site,
and $\mathbf{h}_n=(h_n^x,h_n^y,h_n^z)$ describes a random
field with coefficients drawn independently from a uniform distribution over $[- W, W]$. 

The results in the figure are averaged over $1000$ realizations of the disorder.
They are compared to the random-matrix models \eqref{eq:model} and
\eqref{eq:model2} with patch size $N_0=1$ (corresponding to individual spins,
hence allowing us to reach small localization lengths) and selected values of
$N_\alpha$, also averaged over $1000$ realizations. \red{The values of
$N_\alpha$ are chosen by matching the random-matrix models to the spin chain
result at the equal bipartition (center of the curve), and kept fixed for the
other partition points.} In panel (a) we only take states in the middle of the
spectrum (central 10\% of states in each realization) and vary the disorder
strength, while in panels (b) we fix the disorder strength and vary the energy
range (separating the states in each realization by energy into 10 groups, each
containing 10\% of the states). In all cases, the entropy varies consistently
with choice of the partition point, disorder strength, and energy range, and is
in excellent agreement with the random-matrix models \eqref{eq:model} and
\eqref{eq:model2}. As illustrated in panels (c) and (d), this allows to
determine the effective localization length Eq.~\eqref{eq:xieff} from the data
of the physical model, which is one of the key merits of our approach. \red{This
universal value is independent of the partition.} 
\red{The sudden drop of the effective localization length occurs near $W\approx
5$, which agrees well with the transition point to the many-body localized
behavior in the literature} \cite{PhysRevB.93.174202} \footnote{Note that the
model in Ref.~\cite{PhysRevB.93.174202} is based on spin operators
$\textbf{S}=\boldsymbol{\sigma}/2$, so that the definition of the field $h$
differs from ours by a factor of 2.}. 

\emph{Statistical justification of universality.---}The observed universal
behavior in the random-matrix models \eqref{eq:model} and  \eqref{eq:model2}
follows directly from the statistical properties of the framework. These are
subsumed into two key features, (i) the self-averaging property
\begin{equation}
  \sum_b V^{k|k+1}_{ab}(V_{a'b}^{k|k+1})^*\to \delta_{aa'},
  \label{eq:selfaveraging}
\end{equation}
quickly valid from moderate number of terms in the sum, and (ii) the fact that wave functions drawn from each model constitute a statistically complete basis,
\begin{equation}
  \overline{|\psi\rangle\langle \psi|}=\openone.
  \label{eq:statisticalbasis}
\end{equation}

In particular, in the structured model \eqref{eq:model}, we can use the
self-averaging property to show that for $N_\alpha=1$ the entropy of a partition
$\mathsf{A}|\mathsf{B} = \mathsf{P}_1 \ldots \mathsf{P}_K | \mathsf{P}_{K+1}
\mathsf{P}_{K+2} \ldots$ reduces to that of the patches adjacent to the
partitioning point,
\begin{equation}
  \label{eq:sreduced}
  S(\mathsf{A}|\mathsf{B})\approx
  S(\mathsf{P}_K|\mathsf{P}_{K+1}),
\end{equation}
which in turn is given by Page's result \eqref{eq:page}  for the reduced system.
To see this, let us write the wave function \eqref{eq:model} for $N_\alpha=1$ as
\begin{equation}
\psi_{\mathbf{a}a_Kb_{K+1}\mathbf{b}}=\psi^{(A)}_{\mathbf{a},a_K}V_{a_Kb_{K+1}}\psi^{(\mathsf{B})}_{b_{K+1},\mathbf{b}}
,
\end{equation}
where $a_K$ and $b_{K+1}$ are the indices in the patches next to the partition
point, with Gaussian correlation amplitudes $V\equiv V^{K|K+1}$, while
$\mathbf{a}$ and $\mathbf{b}$ subsume all the other indices (we also drop the
index $\alpha$). The structure implies that the reduced density matrices
$\rho^{\mathsf{A}|\mathsf{B}}$ has the same rank as the truncated density matrix
$\rho^{\mathsf{P}_K|\mathsf{P}_{K+1}}\propto VV^\dagger$: In the space of
indices $\mathbf{a}$, any state orthogonal to the span of
$\psi^{(A)}_{\mathbf{a},a_K}$ corresponds to a vanishing eigenvalue.
Furthermore, to a very good approximation both matrices share the same
entanglement spectrum: The self-averaging property \eqref{eq:selfaveraging}
implies $\sum_{\mathbf{a}}\psi^{(A)}_{\mathbf{a},a_K}\psi^{*(A)}_{\mathbf{a},a_K'}\to\delta_{a_Ka_K'}$. Hence, the finite
eigenvalues of $\rho^{\mathsf{A}|\mathsf{B}}$ are recovered with high accuracy
from approximate eigenvectors $\varphi_{\mathbf{a}a_K}=\psi^{(A)}_{\mathbf{a},a_K}\varphi_{a_K}$ where $\boldsymbol{\varphi}$ is an
eigenvector of $\rho^{\mathsf{P}_K|\mathsf{P}_{K+1}}$. Thereby, the entropy is
given by Page's result for the reduced system of only two adjacent patches, as
stipulated in Eq.~\eqref{eq:sreduced}.

For a finite number $N_\alpha$ of states participating in Eq.~\eqref{eq:model},
a similar selfaveraging argument applies to show that
\begin{equation}
  \label{eq:sreduced2}
  S(\mathsf{P}_1 \ldots \mathsf{P}_K | \mathsf{P}_{K+1} \mathsf{P}_{K+2}
  \ldots) \approx S(\mathsf{R} | \mathsf{R}'),
\end{equation}
with effective ergodic patches $\mathsf{R}$, $\mathsf{R}'$ of increased size
$N_0+\log_2N_\alpha$, hence a reduced system of overall size as given by
Eq.~\eqref{eq:xieff}. Here we use that the collection of states
$\psi^{(A)\alpha}_{\mathbf{a},a_K}$ with reinstated label $\alpha$ remains
statistically orthogonal to each other as long as $N_\alpha$ does not grow too
large. Thereupon, $\rho^{\mathsf{A}|\mathsf{B}}$ shares the entanglement
spectrum of a \emph{direct sum} of matrices
$N_\alpha^{-1}\bigoplus_{\alpha=1}^{N_{\alpha}} V^\alpha V^{\alpha\dagger}$.
Accounting also for
the indicated overall normalization, the entropy then increases by $\log_2N_\alpha$, resulting in Eq.~\eqref{eq:sreduced2}.

Establishing the ergodic behavior for large $N_\alpha$ is equally
straightforward. Even if matrices $V^{\alpha}$ were drawn from a highly
structured distribution, the Wishart ensemble underlying Page's result is
recovered from $V=\sum_\alpha V^{\alpha}$ for large $N_{\alpha}$ as long as the
matrices form a complete basis in a statistical sense. This applies, in
particular also to independently drawn states of the form \eqref{eq:model},
whose span covers the whole space according to their ensemble average
\eqref{eq:statisticalbasis}. Adding a large number $N_\alpha=O(M)=O(2^N)$ of
these states therefore recovers the ergodic case. This expectation is again
compatible with the logarithmic growth of the effective localization length
\eqref{eq:xieff} stipulated above.

For the unstructured model, the same arguments can be adapted in a simplified
form. For $N_\alpha=1$, the entropy vanishes, corresponding to $\xi_0=0$. For
moderate values of $N_\alpha$, we observe the same statistical direct sum as
above, giving rise to the logarithmic scaling, until one reaches the ergodic
limit where we can again utilize that states drawn from Eq.~\eqref{eq:model2}
form a statistically complete basis.

\emph{Relation to matrix-product states.---}To further illuminate our framework
we connect it to the general framework of matrix-product states. This framework
provides a universal representation of many-body states, and arises
mathematically from successive applications of singular-value
decompositions \cite{Schollwoeck2011}---which are also at the heart of the
analysis of the Wishart ensemble \cite{Zyczkowski2001,Majumdar2011}. Starting
from Eq.~\eqref{eq:model}, we can make this connection explicit by using a
singular-value decomposition of the correlation matrices,
\begin{equation}
  V^{k|k+1,\alpha}_{ab} = \sum_l u^{k,\alpha}_{al} \lambda^{k|k+1,\alpha}_l
  v^{k+1,\alpha}_{lb},
\end{equation}
with unitary matrices $u^{K,\alpha}$ and $v^{K,\alpha}$ and diagonal matrices
$\lambda$ of Schmidt coefficients (encoding the entanglement spectrum of the
reduced system with patches $P_K$ and $P_{K+1}$). Rearranging terms and
reinterpreting indices, this directly delivers the representation
\begin{equation}
  \label{eq:mps}
  \psi_{abcd\ldots} = \mathrm{tr}\,(\Gamma^{1,a} \Lambda^{1|2} \Gamma^{2,b}
  \Lambda^{2|3} \ldots),
\end{equation}
with block matrices $\Gamma^{k,a}_{lm}=\bigoplus_\alpha v^{k,\alpha}_{la}
u^{k,\alpha}_{am}$ and $\Lambda^{k|k+1}=\bigoplus_\alpha
\lambda^{k|k+1,\alpha}$. Equation \eqref{eq:mps} is analogous to Vidal's
representation of matrix-product states \cite{Vidal2003}, which naturally
incorporates entanglement characteristics in the diagonal matrices $\Lambda$,
but with two notable practical differences. (i) In Vidal's representation, the
matrices $\Lambda^{k|k+1}$ contain the exact entanglement spectrum of the
bipartition $\mathsf{A}|\mathsf{B}$ of the complete system, while here they
contain the approximate entanglement spectra of the neighboring patches.
However, both spectra accurately agree according to our derivation of
Eq.~\eqref{eq:sreduced}. In spirit, this conforms to the conventional reduction
of matrix-product states by truncation of the entanglement spectrum, hence the
rank of matrices $\Lambda$~\footnote{In Vidal's representation the matrices
$\Gamma$ obey a specific normalization condition, which in our model is observed
upon self-averaging.}. (ii) Canonically, Vidal's representation is designed to
describe a single state, while the block structure above implies this
representation being carried out for each individual state in the sum over
$\alpha$. Thereby, we here also encounter a truncation of the matrices $\Gamma$,
which are of rank $N_\alpha$.

In principle, our framework can therefore also be formulated in the language of
matrix-product states. Arguably, however, the two modifications outlined above
cannot be easily anticipated without the guidance of the physically transparent
form \eqref{eq:model} of the underlying wave function. As shown above, it is
indeed the interplay of these two truncations which sets the universal
entanglement length scale \eqref{eq:xieff}.

\emph{Conclusions.---}In summary, we  proposed a random-matrix framework for
many-body quantum systems that captures the effect of finitely ranged
entanglement, subsumed into a  universal effective entanglement localization
length. The framework allows us to make predictions for the entanglement entropy
for different choice of partitions, which agree well with those of physical
systems with different disorder strengths and energy densities. This provides a
route to extract this effective entanglement localization length from data.

Just as Page's law can be utilized as a benchmark to detect deviations from
ergodic behavior, the models presented here can serve as a useful benchmark to
test concrete hypotheses about disordered many-body systems. For instance, while
the models do not \emph{predict} an entanglement localization transition in
sufficiently disordered system, or discriminate such a transition from a
crossover, they can be employed to investigate this issue based on the described
extraction of the effective localization length.

More broadly, our framework incorporates a form of one-parameter scaling, and
hence also allows to test this as a hypothesis and detect possible deviations.
In particular, in the structured variant of the model the effective localization
length denotes a contiguous ergodic region, while the spatial structure of the
ergodic region is not prescribed in the unstructured model\red{, emphasizing the
universality of this aspect over a range of possible microscopic models}. The
observed universality can be conjectured to extend to interpolating scenarios,
including multifractal scenarios for which no simple model exists. This connects
our approach directly to a crucial question in the analysis of the many-body
localization transition \cite{Potter2015, Vosk2015, Zhang2016, Dumitrescu2017,
Thiery2018, Goremykina2019}, which can be further pursued by considering
entanglement in disjoint  partitions of the system \cite{Herviou2019}.



\begin{acknowledgments}
This research was funded by UK Engineering and Physical Sciences Research
Council (EPSRC) via Grant No. EP/P010180/1. Computer time was provided by
Lancaster University's High-End Computing facility.
\end{acknowledgments}


%

\end{document}